\documentclass[aps,twocolumn,pra,showpacs,floatfix]{revtex4}
\usepackage{epsfig}
\usepackage{graphicx}
\usepackage{dcolumn}
\usepackage{amssymb,amsmath}
\usepackage{mathrsfs}
\begin{document}

\title{Additional clock transitions in neutral ytterbium bring new possibilities for testing physics beyond the Standard Model}

\author{V. A. Dzuba$^1$, V. V. Flambaum$^1$, and S. Schiller$^2$}

\affiliation{$^1$School of Physics, University of New South Wales, Sydney 2052, Australia}
\affiliation{$^2$Institut f\"{u}r Experimentalphysik, Heinrich-Heine-Universit\"{a}t D\"{u}sseldorf, 40225 
D\"{u}sseldorf, Germany}

\begin{abstract}
We study the prospects of using transitions from the ytterbium ground state to metastable states 
$^3{\rm P}^{\rm o}_2$ at 
$E=19\,710.388~$cm$^{-1}$ and $4f^{13}5d6s^2\,(J=2)$ at $E=23\,188.518~$cm$^{-1}$ as clock transitions
in an optical lattice clock. Having more than one clock transition in Yb 
could benefit the search for new physics beyond the Standard Model via studying
the non-linearity of King's plot or the time-variation of the ratio of the frequencies of two clock transitions. 
We calculate the lifetime of the states, relevant transition amplitudes,
systematic shifts, and the sensitivities of the clock transitions to a variation of the fine structure constant 
and to the gravitational potential. 
We find that both 
transitions can probably support 
ultra-high accuracy, similar to what is already 
achieved for the $^1$S$_0$ -  $^3$P$^{\rm o}_0$ clock transition. 
\end{abstract}

\pacs{31.15.A-,11.30.Er}

\maketitle


\section{Introduction.}

The search for new physics beyond  the Standard Model (SM) with low-energy experiments requires 
extremely accurate measurements. 
The highest fractional accuracy has been achieved for atomic optical clocks such as based on
Sr, Yb, Al$^+$, Hg, Hg$^+$, and Yb$^+$. 
It is now at the low-$10^{-18}$ level~\cite{Ludlow,Chou,Beloy1,Beloy2,Ushijima,Nicholson,Katori}. 
When clock transitions are also sensitive 
to new physics beyond the SM, the benefit of using atomic clocks is enormous. 
At least two clock transitions must be available in order to produce a test of the SM.
The clock transitions must have different sensitivity to new physics to avoid cancellation of the 
effect in the ratio of frequencies. A good example of such a system is the Yb$^+$ ion. It has two clock transitions, one is an
electric quadrupole (E2) transition between the ground [Xe]$4f^{14}6s \ ^2$S$_{1/2}$ and the 
excited [Xe]$4f^{14}5d \ ^2$D$_{3/2}$ states,
another is an electric octupole (E3) transition between the ground and the excited [Xe]$4f^{13}6s^2 \ ^2$F$_{7/2}$ states.
This second transition has high sensitivity to a variation of the fine structure constant~\cite{Yb+alpha} and 
to local Lorentz invariance violation~\cite{Yb+LLI}, while the first transition can serve as an anchor. 

It has been recently suggested that the search for a possible non-linearity of King's plot can be used in the search for
new physics beyond the SM~\cite{Kingsplot}. On a King's plot the ratios of isotope shifts for
two atomic transition frequencies are plotted for several isotopes. In the absence of new physics all points are expected
to be approximately on the same line (up to some small corrections~\cite{Viatkina}). Electron-nucleus interaction
due to exchange of a hypothetical scalar particle produces a non-linearity of King's plot.
 The minimum data needed to search for non-linearity of King's plot requires two transitions
and four isotopes (leading to three isotope shifts against a reference isotope). The expected smallness
of the effect suggests the use of clock transitions. 

The ytterbium atom has seven stable
isotopes, one well-studied clock transition, and several metastable states which can probably be used in 
additional clock transitions. In this paper we study the [Xe]$4f^{14}6s6p \ ^3$P$^{\rm o}_2$ and [Xe]$4f^{13}5d6s^2 \ (7/2,3/2)_2$ 
states.
The numbers in parentheses are the angular momenta of the $f$-shell hole and of the $d$-electron. 
The subscript denotes the total electronic angular momentum $J=2$.

\begin{figure}[tb]
\epsfig{figure=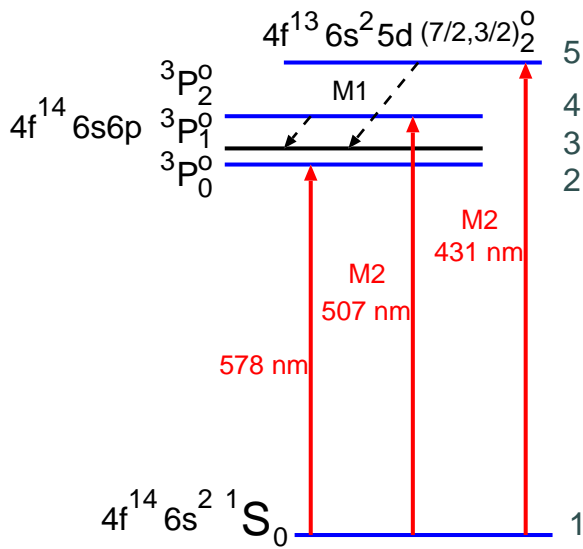,scale=0.9}
\caption{Energy diagram for five lowest states of Yb (approximately to scale). 
Numeration of the states, presented on the right, is the same as in Table~\ref{t:yb}.
There are three metastable states and three clock transitions (1 - 2, 1 - 4 and 1 - 5).
Dominating decay channels for clock states 4 and 5 are M1 transitions to the lower-lying 
$^3$P$^{\rm o}_1$ state.}
\label{f:EL}
\end{figure}

The energy diagram for five lowest states of Yb is presented on Fig.\,\ref{f:EL}. There are three metastable states
and three transitions between ground and metastable states which can be used as clock transitions. 
The first (578\,nm) transition is already used as clock transition in a number of laboratories around the world. In this work we study 
the other two clock transitions.
The transition denoted by 1 - 4 in the follwoing was first observed by Yamaguchi et al.\,\cite{Yamaguchi2010}, and has since been studied in the context of photoassociation and atom-atom interaction physics \cite{Hara2014,Taie2016,Kato2016}. Transition linewidths 
in the kHz-range have been realized \cite{Kato2013}. 
The 1 - 5 transition has not been studied experimentally yet, to the best of our knowledge.

\section{Calculations.}

\begin{table}
\caption{\label{t:yb}
Energies and lifetimes of low-lying states of Yb. New clock states are shown in bold.}
\begin{ruledtabular}
\begin{tabular}{cllc rrr}
\multicolumn{4}{c}{State}&
\multicolumn{2}{c}{Energy [cm$^{-1}$]} &
\multicolumn{1}{c}{Lifetime} \\
\multicolumn{1}{c}{$N$}&
&&\multicolumn{1}{c}{$J$}& 
\multicolumn{1}{c}{Expt.~\cite{NIST}}& 
\multicolumn{1}{c}{CIPT} & \\
\hline
1 & $4f^{14}6s^2$  & $^1$S      &   0 &    0 &      0 & $\infty$ \\

2\footnotemark[1] & $4f^{14}6s6p$  & $^3$P$^{\rm o}$    & 0 & 17288 & 17265 & 23-26\footnotemark[2] s \\
3 &         $4f^{14}6s6p$  & $^3$P$^{\rm o}$                      & 1 & 17992 & 18327 & 500 ns \\
{\bf 4} &  $ {\bf 4f^{14}6s6p}$  & {\bf $^3$P$^{\rm o}$}  & {\bf 2} & {\bf 19710} & {\bf 19895} &  {\bf 15\footnotemark[3] s} \\

{\bf 5} &  ${\bf 4f^{13}5d6s^2}$ & {\bf (7/2,3/2)$^{\rm o}$}& {\bf 2} & {\bf 23188}  & {\bf 24831} &  {\bf 200\footnotemark[3] s} \\
6 &  $ 4f^{13}5d6s^2$ &  (7/2,3/2)$^{\rm o}$&  3 &  27445  &  27185 &   \\

7 & $4f^{14}5d6s$  &  $^3$D           & 1 & 24489  &  27584 &  \\ 
8 &  $4f^{14}5d6s$  &  $^3$D          & 2 & 24751  & 27678  &  \\ 
9 &  $4f^{14}5d6s$  &  $^3$D          & 3 & 25271  & 27763  &  \\ 

10 & $4f^{14}6s6p$ &  $^1$P$^{\rm o}$     & 1 &  25068 & 24753 &  5 ns\\
11 & $4f^{14}5d6s$  &  $^1$D        & 2 &  27677 & 28156 &  \\

12 & $4f^{13}5d6s^2$& (7/2,5/2)$^{\rm o}$ & 1 &  28857 & 29610 &  8 ns \\

\end{tabular}
\footnotetext[1]{Current upper clock state.}
\footnotetext[2]{23.0 s for $^{171}$Yb and 26.0 s for $^{173}$Yb~\cite{quenching1}.}
\footnotetext[3]{For even isotopes.}
\end{ruledtabular}
\end{table}

We perform atomic structure calculations with  the CIPT method~\cite{CIPT}. 
It combines configuration interaction (CI) with the perturbation theory (PT) by treating excited configurations 
perturbatively rather
than including them into the CI matrix. This reduces the size of the CI matrix for the many-electron problem 
by many orders of magnitude, making it possible to deal with systems having a large number of electrons
outside closed shells. This is important for the current problem because we are dealing with 
states of ytterbium which have excitations from the $4f$ subshell. This means that the total number 
of external electrons is sixteen; e.g., in the excited $4f^{13}5d6s^2 \ (7/2,3/2)^{\rm o}_2$ state we have thirteen 
$4f$ electrons, one $5d$ electron and two $6s$ electrons.

The results for the energies of relevant low-energy  states of Yb are presented in Table~\ref{t:yb}. Note some small
differences in the results compared to previous calculations~\cite{CIPT}. 
This is due to differences in the basis and the size the effective CI matrix. These differences are
within the accuracy of the method.

To calculate transition amplitudes we need to include the interaction of the atom with an external  electromagnetic field.
We consider the interaction  in dipole and quadrupole approximation leading 
to electric and magnetic dipole (E1 and M1) and electric and magnetic quadrupole (E2 and M2) transitions.
We use the random-phase approximation (RPA) for the interaction.
The RPA equations for single-electron atomic states have the form
\begin{equation}
\label{eq:RPA} 
(H^{\rm HF} - \epsilon_c)\delta \psi_c = -(\hat F +\delta V^F),
\end{equation}
where $H^{\rm HF}$ is the Hartree-Fock Hamiltonian, the index $c$ enumerates states in the atomic core,
$\delta \psi_c$ is a correction to the core state due to an external field, $\hat F$ is the operator of 
the external field, and $\delta V^F$ is the correction to the self-consistent Hartree-Fock potential
due to the external field. Equations (\ref{eq:RPA}) are solved self-consistently for all states in the core,
leading to the effective
operator of the external field, $\hat F \rightarrow \hat F +\delta V^F$. Matrix elements
between states containing external electrons are calculated by the formula
\begin{equation}
\label{eq:ME} 
A_{ab} = \langle b||\hat F +\delta V^F|| a \rangle,
\end{equation}
where $| a \rangle$ and $| b \rangle$ are many-electron (sixteen for Yb) wave functions found in the CIPT calculations.

We consider interaction of atomic electrons with external field
in dipole and quadrupole approximation leading 
to electric and magnetic dipole (E1 and M1) and electric and magnetic quadrupole (E2 and M2) transitions.

The rates of spontaneous emission are given in atomic units by
\begin{equation}
\label{eq:E1} 
T_{E1,M1} = \frac{4}{3}(\alpha\omega)^3\frac{A_{E1,M1}^2}{2J+1},
\end{equation}
for electric dipole (E1) and magnetic dipole (M1) transitions, and by
\begin{equation}
\label{eq:E2} 
T_{E2,M2} = \frac{1}{15}(\alpha\omega)^5\frac{A_{E2,M2}^2}{2J+1},
\end{equation}
for electric quadrupole (E2) and magnetic quadrupole (M2) transitions. In these formulae
$\alpha$ is the fine structure constant,
$\omega$ is the frequency 
(not angular frequency)
of the transition in atomic units, $A$ is the amplitude of the transition in atomic units, and $J$
is the total angular momentum of the upper state. The magnetic amplitudes $A_{M1}$ and  $A_{M2}$
are proportional to the Bohr magneton 
$\mu_B = |e|\hbar/2mc$. Its value in Gaussian-based atomic units is
$\mu = \alpha/2 \approx 3.6\times 10^{-3}$.

The calculated amplitudes and corresponding transition rates are presented in Table~II.
Note that the value of the electric dipole transition amplitude between states number 1 (ground state) and
state number 10 (which is one of the transitions used for laser cooling of the Yb atom), $\langle 1||E1||10\rangle = 4.14$~a.u., 
is in excellent agreement with the experimental
value of 4.148(2)~a.u.~\cite{Yb-E1}. This is in sharp contrast to the large disagreement between experiment
and the value of 4.825 a.u. obtained in very sophisticated calculations treating the Yb atom as a 
two-valence-electron system~\cite{YbDD}. 
The reason for this disagreement is the strong mixing between the $4f^{14}6s6p \ ^1$P$^{\rm o}_1$
and  $4f^{13}5d6s^2 \ (7/2,5/2)^{\rm o}_1$ states (states number 10 and 12 in Table~I). This mixing 
cannot be properly taken into account in the two-valence-electrons approximation.
See Ref.~\cite{YbDD} for a detailed discussion. In current CIPT calculations we explicitly include mixing
between three odd configurations, $4f^{14}6s6p$, $4f^{14}5d6p$, and $4f^{13}5d6s^2$, while all other
configurations are included perturbatively. 

In even isotopes the $^1$S$_0$ -  $^3$P$^{\rm o}_0$ clock transition is extremely weak but can be 
opened by a magnetic field.
The electric dipole amplitude between two states $a$ and $b$ induced by a magnetic field is
\begin{eqnarray}
\label{eq:AH} 
A_{\rm B,ab} = \left(\sum_n \frac{\langle b|M1| n \rangle \langle n|E1| a \rangle}{E_b-E_n} + \right. \nonumber \\
 \left. \sum_n \frac{\langle b|E1| n \rangle \langle n|M1| a \rangle}{E_a-E_n} \right) \times B .
\end{eqnarray}
Here $B$ is the magnetic field directed along the $z$ axis. The $z$ components of the electric dipole and magnetic dipole
matrix elements are used; the summation goes over the complete set of intermediate states. 
In the SI system, the atomic unit for magnetic field is
1~a.u.$=2.35\times 10^5$~T.  The transition rate is given by (\ref{eq:E1}) while the 
angular frequency of the Rabi oscillations is 
$\Omega_R =  E_0 A_{\rm B,ab}$. Here $E_0$ is the amplitude of the laser electric field.
Considering the $^1$S$_0$ - $^3$P$^{\rm o}_0$ (1 - 2) transition and using amplitudes from Table~II for the three 
first contributions to eq.\,(\ref{eq:AH}) we find that in SI units 
$\Omega_R/2\pi= (242~$Hz/${\rm T}\sqrt{{\rm mW/cm}^2})B\sqrt{I}$. Here, $I$ is the intensity of the laser wave.
The coupling coefficient is in good agreement with the value
$186~$Hz/${\rm T}\sqrt{{\rm mW/cm}^2}$ from Ref.~\cite{Taich}.

\begin{table}
\caption{\label{t:tr}
Transition amplitudes ($A$, in atomic units), corresponding rates ($R$) of spontaneous emission and experimental 
transition frequencies between some states of Table~\ref{t:yb}. Numbers in square brackets indicate powers of ten. 
New clock transitions are shown in bold.
To obtain $A_{\rm E2}$, $A_{\rm M2}$ in SI units, multiply by  $e\,a_0^2$.}
\begin{ruledtabular}
\begin{tabular}{c clrl}
\multicolumn{1}{c}{Transition}&
\multicolumn{1}{c}{Type}&
\multicolumn{1}{c}{$A$}&
\multicolumn{1}{c}{$\omega$ [cm$^{-1}$]} &
\multicolumn{1}{c}{$T$ [s$^{-1}$]}\\ 
\hline
3  -   1 & E1 & 0.711 & 17992 & 2.0[+6]\\
10  -   1 & E1 & 4.14  &  25068 & 1.8[+8]\\
12  -  1 & E1 & 2.71 &  28857 & 1.2[+8]\\
{\bf 4 - 1} & {\bf M2} & {\bf 0.61[-1]} & {\bf 19710} & {\bf 2.5[-4]}\\
4 - 3 & M1 & 0.57[-2] & 1718 & 6.7[-2] \\
4 - 5 & M1 & 0.37[-4] & 3478 &  \\
{\bf 5 - 1} & {\bf M2} & {\bf 0.993[-2]} & {\bf 23188} & {\bf 1.5[-5]}\\
5 - 2 & E2  &  1.43       & 5900 & 3.3[-4]\\
5 - 3 & M1 & 0.277[-3] &  5196 & 4.7[-3]\\
5 - 4 & M1 & 0.370[-4] & 3478 & 2.6[-5]\\
5 - 6 & M1 & 0.25[-2] & 4257 & \\
5 - 12 & M1 & 0.49[-2] & 5669 & \\
7 - 5 & E1 & 0.531[-1] & 1301 & \\
8 - 5 & E1 & 0.186      & 1563 & \\
9 - 5 & E1 & 0.453      &  2086 & \\
11 - 5 & E1 & 0.744[-1] & 4489 & \\
\end{tabular}
\end{ruledtabular}
\end{table}

\section{Analysis}

The calculated amplitudes and corresponding transition rates are presented in Table~\ref{t:tr}.
 Using data from the Table we find that the lifetimes of the new clock states (number 4 and 5
in Table~\ref{t:yb}) are
about 15\,s and 200\,s respectively. This leads to the quality factors 
$Q = 2\pi\omega/R \approx 10^{16}$ and $10^{17}$. 
The decay of these states is dominated by the M1 transitions to the $^3$P$^{\rm o}_1$ state. 
The rates for the M2 transitions between  
these clock states and the ground state are $2.5 \times 10^{-4}$ s$^{-1}$ and $1.5 \times 10^{-5}$ s$^{-1}$, respectively. 
They are smaller than the rate of the hyperfine interaction-induced 
transition between
clock state 2 ($^3$P$^{\rm o}_0$) and the ground state, which varies between 
$10^{-2}$ and $10^{-1}$ s$^{-1}$ depending on the isotope and the hyperfine structure components~\cite{quenching1}. 
For comparison, they are larger than the rate of  the hyperfine interaction-induced
 E3 transition in Yb$^{+}$ ions, which is $\sim 10^{-6}$ s$^{-1}$~\cite{quenching2}.

\subsection{Rabi oscillations.}

In this paper we focus on even Yb isotopes to avoid large Zeeman shifts 
 in M2 clock transitions (see below). 

The Rabi frequency of a M2 transition is given by $\Omega_{\rm R} = 2 E_0\,\alpha\,\omega\, A_{M2}$,
(all values are in atomic units)
Using the values of the M2 transition amplitudes from Table~\ref{t:tr} we find
$\Omega_R/2\pi= (88,\,17)~$Hz$\sqrt{I}/\sqrt{{\rm mW/cm}^2}$ for the 
1 - 4 and 1 - 5 clock transition, respectively.
The laser intensity $I_{\pi}$ required to achieve a desired excitation time $T_{\pi}=\pi/\Omega_R\simeq1\,{\rm s}$ 
is of order $\mu$W/cm$^2$ and lower, a very small value.

\subsection{Polarizabilities, black-body radiation and Stark shifts.}


\begin{table}
\caption{\label{t:pol0} 
Static scalar  ($\alpha^{\rm S}$) and and tensor ($\alpha^{\rm T}$) polarizabilities
for Yb clock states, in atomic units. Numeration of the states is from Table~\ref{t:yb}.} 
\begin{ruledtabular}
\begin{tabular}{ccc}
\multicolumn{1}{c}{State}&
\multicolumn{1}{c}{$\alpha^{\rm S}(0) $}&
\multicolumn{1}{c}{$\alpha^{\rm T}(0) $}\\
\hline
1 & 150 & 0 \\
2 &  304  & 0 \\
4 & 418 & -70 \\
5 & 124 & -6 \\
\end{tabular}
\end{ruledtabular}
\end{table}

The shifts of the clock frequency due to the effect of black-body radiation (BBR), of the lattice laser field that 
traps the atoms, and of the clock laser field depend on the  dipole polarizabilities of the clock states. 

The total dipole polarizability of a state  with angular momentum $J \geq 1$ in a laser field of frequency 
$\omega_L$, linearly polarized 
and parallel to the quantization direction is \cite{Derevianko2011}
\begin{equation}
\label{eq:pol} 
\alpha(\omega_L) = \alpha^{\rm S}(\omega_L) + \frac{3J_z^2 - J(J+1)}{J(2J-1)}\alpha^{\rm T}(\omega_L)\,,
\end{equation}
where $\alpha^{\rm S}$ and $\alpha^{\rm T}$ are the dynamic scalar and tensor dipole polarizabilities, 
respectively and $J_z$ is the projection of $J$. 
More general polarization geometries are treated later in section~\ref{s:lightshifts}.
States 1 and 2 have $J=0$ 
(or, in case of fermionic isotopes with nuclear spin 1/2, $F=1/2$), 
thus $\alpha^{\rm T} \equiv 0$. For states 4 and 5, $J=2$, and $\alpha$ depends on $J_z$:
\begin{eqnarray}
 \alpha_a = \alpha^{\rm S} - \alpha^{\rm T},     &{\rm for}&  J_z=0, \label{eq:jz0} \\
 \alpha_a = \alpha^{\rm S} - \alpha^{\rm T}/2, &{\rm for}&  J_z=\pm 1, \label{eq:jz1} \\
 \alpha_a = \alpha^{\rm S} + \alpha^{\rm T},   &{\rm for}&  J_z=\pm 2. \label{eq:jz2}
\end{eqnarray}
The 
polarizabilities  of an atomic state $a$ 
with angular momentum $J$ 
are given by
\begin{eqnarray}
&&\alpha^{\rm S}_a(\omega_L)  = \frac{2}{3(2J+1)} \sum_n \frac{(E_n-E_a)\langle a || \hat D || n \rangle^2}{(E_n-E_a)^2-\omega_L^2},     
\label{eq:alpha0}\\
&&\alpha^{\rm T}_a(\omega_L) = 2\sqrt{\frac{10J(2J-1)}{3(2J+3)(2J+1)(J+1)}} \times \label{eq:alpha2} \\
&&\sum_n (-1)^{J+J_n} \left\{ \begin{array}{lll} 1 & 1 & 2 \\ J & J& J_n \end{array} \right\} 
\frac{(E_n-E_a)\langle a || \hat D || n \rangle^2}{(E_n-E_a)^2-\omega_L^2},   \nonumber  
\end{eqnarray}
where the summation goes over the complete set of 
excited many-electron states, $\hat D$ is the operator of the electric dipole interaction in the valence space,
$\hat D = \sum_v (d_v + \delta V_v)$. Here $d_v=-er_v$, $\delta V_v$ is the RPA correction to the electric dipole 
operator acting on electron $v$ (see eqs.\,(\ref{eq:RPA},\ref{eq:ME})), and the summation over $v$ is the summation 
over the valence electrons. There are also core and core-valence contributions to the scalar polarizability.
We calculate them as described in Ref.~\cite{DF-HI2016}.

The polarizabilities are well-known for the ground and the $^3$P$^{\rm o}_0$ states (see, e.g.~\cite{YbDD,Safronova}). 
There are also experimental and theoretical studies of the polarizabilities
of state 4~\cite{Yamaguchi,Yamaguchi2010,Khramov14}. However, we are not aware of any calculations or
measurements for state 5. We perform the calculations for all clock states, 1, 2, 4 and 5 using two different 
approaches. States 1, 2 and  4 are the states with two valence electrons 
above the closed $4f$ shell. Therefore, we apply a well-developed techniques to perform the calculations
(see, e.g.~\cite{YbDD,DF-HI2016}). 

State 5 has a hole in the $4f$ subshell and requires a different treatment. In principle, one can directly use
expressions (\ref{eq:alpha0}) and (\ref{eq:alpha2}), substituting experimental energies and calculated 
electric dipole matrix elements. This is useful at least for checking the contributions of low-lying resonances. 
However, it does not give correct polarizability due to a large contribution of the highly excited states, 
e.g.  of the $4f^{13}5d6s6p$ configuration. Inclusion of highly excited states in the CIPT method is 
computationally very expensive. Therefore, in addition to direct summation, we use an approach developed 
in Ref.~\cite{AKozlov} for atoms with open $f$-shells. It uses the fact that polarizabilities of such atoms are 
dominated by matrix elements between states with the same $4f^n$ or $5f^n$
subshell, i.e. excitations from the $f$-shell can be ignored. In our case this means that the $4f^{13}$ subshell is
treated as a closed shell with occupation number 13/14. Then the remaining valence electrons form the 
$6s^25d \ ^2$D$_{3/2}$ state similar to the ground state of Lu. The polarizabilities are calculated as for 
a three--valence-electron system having the $6s^25d \ ^2$D$_{3/2}$ ground state.
This approach  gives reasonably good results at least at some distance from resonances~\cite{AKozlov}. 
We do not calculate the polarizabilities of state 5 beyond first resonance (at $\omega \sim 0.04$~a.u.)
because closeness to resonances makes calculations in this region unreliable. 

The results are presented in  Tables~\ref{t:pol0} and \ref{t:trans} and further discussed in section~\ref{s:lightshifts}. 
Static polarizabilities ($\omega_L=0$) are presented in Table~\ref{t:pol0}.
"Magic" frequencies occur when the polarizabilities of the two clock states are equal so that the frequency 
of the transition is not sensitive the electric field strength of the lattice wave. The magic frequencies are
discussed in section~\ref{s:lightshifts}.

Earlier polarizability calculations were performed in refs.~\cite{YbDD,Safronova,Yamaguchi,Yamaguchi2010,Khramov14,Hara2014}.
In particular, Khramov et al.\,\cite{Khramov14} predicted magic wavelengths for the 1 - 4 transition based on calculated polarizabilities.

Our results for states 1, 2 and 4 are 
in good agreement with the earlier calculations and 
and with available experimental data. 
E.g., our value for the difference of static polarizabilities of states 1 and 2,
$\Delta \alpha(0)$ = 154~a.u. (see Table~\ref{t:pol0}) differs by less than 6\% from the experimental values
$\Delta \alpha(0)$ = 146.1(1.3)~a.u.~\cite{Beloy2} and $\Delta \alpha(0)$ = 145.726(3)~a.u.~\cite{Sherman}.
The ratios of the polarizabilities of state 4 for $J_z=0,-1,-2$ to the 
polarizability $\alpha_1$ of the ground state 1, 
were measured in Ref.~\cite{Khramov14} at the laser wavelength $\lambda$ = 1064~nm. Our calculated values for 
these ratios, $\alpha_4(J_z=0)/\alpha_1=1.35$,  $\alpha_4(J_z=-1)/\alpha_1 =1.06$,  
$\alpha_4(J_z=-2)/\alpha_1 =0.20$,
agree well with the experimental values 1.6(2), 1.04(6), and 0.20(2)~\cite{Khramov14}.


\begin{table}
\caption{\label{t:trans} Computed polarizabilities at the clock transition frequencies, differential polarizabilities 
$\Delta\alpha$ and corresponding frequency shift coefficients  due to intensity of the probe laser. It is assumed
for states 4 and 5 that 
$\alpha_a=\alpha_a^{\rm S}-\alpha_a^{\rm T}$.}
\begin{ruledtabular}
\begin{tabular}{cccccc}
\multicolumn{1}{c}{Transition}&
\multicolumn{1}{c}{$\omega_L$}&
\multicolumn{1}{c}{$\alpha_1(\omega_L)$}&
\multicolumn{1}{c}{$\alpha_a(\omega_L)$}&
\multicolumn{1}{c}{$\Delta\alpha_{a1}(\omega_L)$}&
\multicolumn{1}{c}{Stark shift}\\
\multicolumn{1}{c}{1 - $a$}&
\multicolumn{1}{c}{(cm$^{-1}$)}&
\multicolumn{1}{c}{(a.u)}&
\multicolumn{1}{c}{(a.u)}&
\multicolumn{1}{c}{(a.u)}&
\multicolumn{1}{c}{(Hz\,cm$^2$/W)}\\
\hline
1 - 2 &  17288 & 315 & 6 & -310 & ~~15 \\
1 - 4 &  19710 & 355 & $< 10^3$ & $<10^3$ & $<10^2$ \\
1 - 5 &  23189 & 961 & $< 10^3$ & $< 10^3$ & $< 10^2$ \\
\end{tabular}
\end{ruledtabular}
\end{table}

For the BBR shift it is sufficient to consider the difference in polarizabilities $\Delta\alpha_{ab}(0)$ of the 
two clock states at zero frequency.
For the 1 - 4 clock transition $\Delta\alpha_{41}(0)=268$~a.u. (see Table~\ref{t:pol0}).
This is  approximately 1.7 times larger than for the 1 - 2 transition.
Correspondingly, the BBR shift is also 1.7 times larger, i.e. 
$\Delta \omega_{\rm BBR,14}\simeq \ 1.7$~Hz at 300\,K~\cite{YbDD}.  
In contrast, the BBR shift of the 1 - 5 clock transition is about 6 times smaller,
$\Delta \omega_{\rm BBR,15}\simeq \ 0.2$~Hz at 300\,K.

We may thus expect that for both new clock transitions the BBR shifts can be controlled, respectively, at the same 
level as or better than that for the current Yb clock transition 
1 - 2, 
$1\times10^{-18}$ \cite{Beloy2}.

We now estimate the clock transition Stark shift due to the clock 
laser electric field ("probe shift"). It is given by (in atomic units)
\begin{equation}
\Delta\omega_{\rm p}\approx -\Delta\alpha_{ab}(\omega_L) \left(\frac{\varepsilon_{\rm p}}{2}\right)^2,
\label{eq:stark}
\end{equation}
where $\varepsilon_{\rm p}$ is the amplitude of the clock laser electric field. Computed polarizabilities 
and corresponding Stark shifts of the clock transitions are listed in Table~\ref{t:trans}.
The number for the 1 - 2 transition is in exact agreement with the result of Ref.~\cite{Taich}.
In contrast, for the 1 - 4 and 1 - 5 transitions we can only give rough estimations. This is because 
calculations become unreliable when frequency comes close to a resonance. 
At $\omega_L = E_a$ resonance contributions come from states with $E_n \approx 2E_a$ which satisfy
electric dipole selection rules for the $n-a$ transition. There are four states of the $4f^{14}6s5d$ configuration
with energies 39\,808, 39\,838, 39\,966 and 40\,061~cm$^{-1}$ which give dominant contribution to the polarizability
of state 4, and there are four states of the $4f^{13}5d6s6p$ configuration with energies $E_n$=45\,338, 
45\,595, 46\,395, and 46\,431~cm$^{-1}$ which give dominant contribution to the polarizability of state 5.
Assuming that the amplitude is $\sim 1$~a.u. and using Eq.~(\ref{eq:alpha0}) we get 
$\alpha^{\rm S}_5(\omega_L) < 10^3$~a.u. 


Finally, we briefly consider the relevance of static Stark shifts stemming from the electric field produced by undesired stray charges on the windows of the vacuum chamber.
In the field of lattice clocks it is known how to measure these so that the d.c. Stark shift uncertainty is at the $10^{-18}$ 
level for Yb, and also for Sr.
Therefore, we only need to discuss the Stark shift coefficients $\Delta\alpha_{ab}(0)=\alpha_a(0)-\alpha_b(0)$ of the
 proposed transitions and compare them with that of the conventional clock transition. Table\,\ref{t:pol0} shows that 
 $\Delta\alpha_{ab}(0)$ for the 1 - 4 transition is similar to that of the conventional 1 - 2 transition. On the other hand, 
 as mentioned above, for the 1 - 5 transition it is significantly smaller. Thus, there is no critical systematic shift issue here. 

\subsection{Zeeman shifts.}

The clock states considered in this work have the relatively large value of the total electronic angular momentum $J=2$.
This means that they could be sensitive to external magnetic field and electric field gradients. 
We can consider fermionic and bosonic Yb atoms. 

With fermionic isotopes, the total atomic angular momentum $F$ will be half-integer and thus there will be no states 
with zero total spin projection $F_z$.
For lattice clocks with fermionic isotopes, it is a standard experimental practice to cancel the first-order Zeeman shift 
by alternatingly probing two transitions with opposite values of $F_{z,a}-F_{z,b}$ and averaging the two transition frequencies.
However, the Zeeman shift in the standard 1 - 2 transition is a nuclear Zeeman shift. In the present transitions 1 - 4, 1 - 5, 
the electronic Zeeman shift occurs, with electronic Land{\'e} factor $\simeq1.5$ and shifts 
$\approx$(10\,GHz/T)$\times F_{z,a}$. 
In order to achieve an uncertainty of the residual first-order shift equal to $1\times10^{-18}$ on either transition, 
the uncertainty of the magnetic field variation between the alternating measurements must be 
$\le6\times10^{-10}$\,G$/F_{z,a}$, not necessarily on the time scale between interrogations (several seconds) 
but over an appropriate averaging time interval. It will be difficult to achieve this, although the availability of several 
Zeeman states $F_z$ and two clock transitions may help to devise appropriate strategies.

All bosonic Yb isotopes, including the radioactive ones with macroscopic liftetimes, have nuclear spin 0. 
Thus, the two clock states 4 and 5 have nonzero $F=J=2$. 
From the point of view of Zeeman shifts, the bosonic isotopes are advantageous, since levels 4, 5 as well as the ground 
state offer $F_z=J_z=0$ states. The first-order Zeeman shift is then absent for transitions between such states. 
Therefore, we shall focus on the bosonic isotopes in the following.
 
The quadratic Zeeman shift can be estimated using second-order perturbation theory,
\begin{equation}
\label{eq:Z2}
\delta E_Z(J,J_z) = \sum_{n}\frac{|\langle n,J_n,J_z | \mu_z H_z | J,J_z \rangle|^2}{E_J-E_{n}},
\end{equation}
where $J_n=J,J \pm 1$ and the summation goes over the complete set of states. In most cases, the summation 
is strongly dominated by terms within the same fine-structure multiplet. This is because of small energy denominator
and large value of magnetic dipole matrix elements. However, for clock state number 5 (see Table~\ref{t:yb}) three
states give significant contribution, states number 6, 10 and 12. The first two are within the same fine structure multiplet
as clock state 5, while state 12 is strongly mixed with state 10.
The Zeeman shifts calculated for the three clock states are presented in Table~\ref{t:Z2}.
There are several things to note:

(1) The largest shift coefficient
is for the $^3$P$^{\rm o}_0$ state. This is due to the small fine structure interval of 704~cm$^{-1}$
between the $^3$P$^{\rm o}_0$ and $^3$P$^{\rm o}_1$ states.

(2) $J_z=0$ states of levels 4 and 5 have shifts smaller than that of the conventional level 2.

(3) For state 4 the quadratic shift is extremely small for $J_z= \pm 2$ because there is no mixing with this state within the 
fine-structure multiplet.  The shift is due to the M1 matrix elements with states of different configurations. Such matrix 
elements are very small due to the orthogonality of the wave functions.
The shift is further suppressed by large energy denominators.

(4) For state number 5 the shift coefficient for $J_z= \pm 2$ is also small. 
This may be convenient if such states are used for particular purposes where suppression of linear Zeeman shift is aimed for.

The quadratic Zeeman shift for the ground state is small, since there are no fine-structure contributions. 
It is much smaller than for the upper clock states and can be neglected in the difference. 


\begin{table}
\caption{\label{t:Z2}
Second-order Zeeman shift  coefficient for clock states, in units Hz/G$^2$. 
$J_n$ denotes the contributing states.
}
\begin{ruledtabular}
\begin{tabular}{ccccccc}
\multicolumn{1}{c}{State}&
\multicolumn{1}{c}{$J$}&
\multicolumn{1}{c}{$J_n$}&
\multicolumn{1}{c}{$J_z$}&
\multicolumn{3}{c}{Shift} \\
& &&&\multicolumn{1}{c}{This work}&
\multicolumn{2}{c}{Other}\\
\hline
2 & 0 & 1 & 0 & -6.0[-2] & -6.2[-2]\footnotemark[1] & -7(1)[-2]\footnotemark[2] \\
4 & 2 & 1,2,3 & 0 & 1.2[-2] &&\\
   &    &    &1 & 9.2[-3] &&\\
   &    &    &2 & -4.7[-7] &&\\
5 & 2 & 1,2,3 &0 & -4.3[-3] && \\
   &    &    &1 & -3.4[-3] &&\\
   &   &     &2 & -3.4[-3] && \\
\end{tabular}
\footnotetext[1]{Theoretical estimation, Ref.~\cite{Taich}.}
\footnotetext[2]{Experiment, Ref.~\cite{Lemke}.}
\end{ruledtabular}
\end{table}

\subsection{Electric quadrupole shift.}

The energy shift due to a gradient of a residual static electric field $\varepsilon$
is described by a corresponding term in the Hamiltonian
\begin{equation}\label{eq:HQ}
\hat H_Q = -\frac{1}{2}\hat Q\frac{\partial \varepsilon_z}{\partial z},
\end{equation}
where $\hat Q$ is the atomic quadrupole moment operator ($\hat Q = |e|r^2 Y_{2m}$, the same as for E2 transitions). 
The energy shift of a state with total angular momentum $J$ is proportional of the atomic quadrupole moment
of this state. It is defined as twice the expectation value of the $\hat Q$ operator 
in the stretched state
\begin{equation}\label{eq:Q}
Q_J = 2\langle J,J_z=J|\hat Q| J,J_z=J \rangle.
\end{equation}
Calculations similar to those described above give the values 
$Q_J=-18$~a.u. for state 4, and $Q_J=-5.3$~a.u. for state 5.
For a state with projection $J_z$ of the total angular momentum $J$, the shift is proportional to $3J_z^2-J(J+1)$. 
Note that if $J=2$ the shift has the same value, but opposite sign, for $J_z=0$ and $J_z=\pm2$. 
Therefore, averaging over these states can, at least in principle, suppress both electric quadrupole and linear Zeeman shifts. In addition, the vector light shift and the tensor light shift cancel (not the scalar).

Alternatively, it is possible to reduce the quadrupole shift by measuring the transition frequency three times, 
with the magnetic field direction applied in three mutually orthogonal directions~\cite{Itano}. 
In this case one can use only states with $J_z=0$ avoiding the linear Zeeman shift. 

\subsection{Search for variation of the fine structure constant.}

\begin{table}
\caption{\label{t:alpha}
Sensitivity of Yb clock transitions to variation of the fine structure constant. 
Transition frequencies are experimental.}
\begin{ruledtabular}
\begin{tabular}{ccrr}
\multicolumn{1}{c}{Clock}&
\multicolumn{1}{c}{Transition frequency} &
\multicolumn{1}{c}{$q$} &
\multicolumn{1}{c}{$K=2q/\omega_0$} \\
\multicolumn{1}{c}{transition} & 
\multicolumn{1}{c}{$\omega_0$ (cm$^{-1}$)} & 
\multicolumn{1}{c}{ (cm$^{-1}$)}  & \\
\hline
1 - 2  &  17288.439 &   2714\footnotemark[1] & 0.31 \\
1 - 4  &  19710.388 &   5505  & 0.56 \\
1 - 5  &  23188.518 & -44290 & -3.82 \\
\end{tabular}
\footnotetext[1]{Ref.~\cite{Yb+alpha}.}
\end{ruledtabular}
\end{table}

To search for a possible time variation of the fine structure constant $\alpha$ one needs to monitor a ratio of two clock
frequencies 
$i,\,j$ over a long period of time. Atomic calculations are needed to link a variation of frequencies to a variation
of $\alpha$. It is convenient to express the atomic transition  frequencies in a form
\begin{equation}
\label{eq:q} 
\omega = \omega_0 + q\left[\left(\frac{\alpha}{\alpha_0}\right)^2-1\right],
\end{equation}
where $\omega_0$ and $\alpha_0$ are present-time values of the frequency and 
the fine structure constant, and $q$ is the sensitivity coefficient which comes from calculations.
Then
\begin{equation}
\label{eq:w} 
\frac{\partial}{\partial t} \ln\frac{\omega_i}{\omega_j} = \frac{\dot \omega_i}{\omega_i} - \frac{\dot \omega_j}{\omega_j} =
\left(\frac{2q_i}{\omega_i} - \frac{2q_j}{\omega_j}\right)\frac{\dot \alpha}{\alpha}.
\end{equation}
To find the values of $q$ for each transition we calculate the frequencies of the transitions at different values of $\alpha$ and
then calculate the derivative numerically. The values of $q$ and corresponding enhancement factors $K=2q/\omega_0$ are
presented in Table~\ref{t:alpha}. 
We see that the 1 - 5 transition is the most sensitive to the variation of the fine structure
constant. If we compare it to the currently used 1 - 2 transition then
\begin{equation}
\label{eq:wq} 
 \frac{\dot \omega_{15}}{\omega_{15}} - \frac{\dot \omega_{12}}{\omega_{12}} = 4.12 \frac{\dot \alpha}{\alpha}.
\end{equation}
In other words, there is significant enhancement of the variation of the frequency ratio
compared to the variation of the fine
structure constant. The enhancement comes from the different nature of the two clock transitions. 
The clock transition 1 - 2 corresponds to the $s-p$ single-electron transition, while the 
clock transition 1 - 5 is the $f-d$ transition.

The 
figure of merit $F_{ij}$ associated with a particular transition pair $i,\,j$
is the ratio of $\alpha$-sensitivity to the (absolute) systematic uncertainty $u$ of the frequency ratio,
\begin{equation}
\label{eq:figure_of_merit} 
F_{ij}= \frac{|K_i-K_j|}{u(\omega_i/\omega_j)} = \frac{|2q_i/\omega_i-2q_j/\omega_j|}{\sqrt{(u(\omega_i)/\omega_i)^2+
(u(\omega_j)/\omega_j)^2}}\,.
\end{equation}
We note that the uncertainties of the transitions, $u(\omega_i)$, $u(\omega_j)$, do not possess any natural proportionality 
to their respective transition frequencies. Therefore, if one of the transition frequencies is significantly smaller than the other, 
no particular advantage results. 
At the present level of analysis of the systematic shifts, it appears that comparing the optical 
transitions 1 - 2 and 1 - 5 would be as performant as the comparison of 1 - 2 and the infrared transition 2 - 5 proposed in 
Ref.\,\cite{SafronovaYb} (see also discussion in Section~\ref{other}).

\subsection{Search for 
	Einstein Equivalence Principle violation.}

Theories 
attempting 
the unification of gravity with other interactions suggest that the Einstein
equivalence principle (EEP) might be violated at high energy~\cite{KS89}. It might be possible to 
discover evidence for the violation 
at low energies by 
observing tiny 
variations of atomic frequencies
in a varying graviational potential.
High-precision atomic clocks can be used to search for such variations~\cite{DF17}.
In the framework of the Standard Model Extension (SME), the term in the hamiltonian responsible for the EEP violation can be presented
in the form (see, e.g.~\cite{CK98,Saf-Nature})
\begin{equation}
\hat H_{c_{00}} = c_{00}\frac{2}{3}\frac{U}{c^2} \hat K,
\label{eq:dH}
\end{equation}
where $c_{00}$ is  one of the  parameters in the SME characterising the magnitude of the EEP violation, 
$U$ is the gravitation potential,
$c$ is the speed of light, 
$\hat K = c\gamma_0 \gamma^jp_j/2$ is the relativistic operator of kinetic energy in which $\gamma^j$ are Dirac matrices, and 
${\bf p}=-i\hbar \nabla$ is  electron momentum operator. 
Upper limits for
$c_{00}$ 
can be determined experimentally by measuring the frequency ratio of two dissimilar, co-located clocks, 
as a function of the local gravitational potential $U$,  
\begin{equation}
\frac{\omega_j}{\omega_i}\Delta\left(\frac{\omega_i}{\omega_j}\right)
=\frac{\Delta \omega_i}{\omega_i} - \frac{\Delta \omega_j}{\omega_j} = 
-(R_i-R_j)\frac{2}{3}c_{00}\frac{\Delta U}{c^2}\ .
\label{e:R1R2}
\end{equation}

$R_i$ are relativistic factors  which describe the deviation of the expectation value of 
the kinetic energy $E_K$ from the value given by the virial theorem (which states $E_K=-E$, where $E$ is the total energy), 
\begin{equation}
R_{ba}=-\frac{E_{K,a} -E_{K,b}}{E_a-E_b}\,.
\label{e:R}
\end{equation}
$\Delta U$ is the change of the gravitational potential
between the measurements. Experimentally, one should make several measurements during at least one year and search 
for a correlation between atomic frequency ratio and the Earth-Sun distance (see~\cite{DF17} for details). 

One needs two atomic transitions with different values of $R$.
Both the size $|R_i-R_j|$ {\it and} the experimental inaccuracies of the determination of the two frequencies 
$\omega_i$, $\omega_j$ are critical parameters determining the sensitivity of the test, 
analogously to the earlier discussion.

The value of $R$ can be 
found from relativistic atomic calculations. For the 1 - 2 clock transition it was calculated in 
Ref.~\cite{DF17},  $R_{12}=1.20$. 
Transitions which are sensitive to a variation of the fine structure constant should have relativistic factors significantly different from the 
non-relativistic limit  $R=1$. We calculated the relativistic 
factors for the 1 - 4 and 1 - 5 clock transitions
using the approach of Ref.~\cite{DF17}  
and the CIPT method. The results are $R_{14}= 1.40$ and $R_{15}=0.62$. These values imply a good sensitivity
to the EEP-violating term in eq.\,(\ref{eq:dH}). For example, if the frequencies of the 1 - 2 and 1 - 5 transitions are compared, then
\begin{equation}
\frac{\Delta \omega_{12}}{\omega_{12}} - \frac{\Delta \omega_{15}}{\omega_{15}} = 
-0.37 c_{00}\frac{\Delta U}{c^2}.
\label{e:EEP}
\end{equation}
This 
value of $|R_i-R_j|$ is higher 
than for most other optical clock transitions 
(with exceptions of Yb$^+$ and Hg$^+$)~\cite{DF17}. 

\subsection{Search for new physics using the non-linearity of King's plot.}

In the King's plot the isotope shift of an atomic transition is plotted against the isotope shift of another transition. 
This is done for several isotopes with every new isotope adding a new point on the plot. Normally, all points are on
the same line. This is a consequence of the factorisation of the nuclear and electron variables in the field (volume)
shift term. See Ref.~\cite{Jin} for the case of Yb.
However, if there is a new interaction between atomic electrons and nucleus which depends on the
number of neutrons, the factorisation and thus linearity might be broken. 
The expected small value of the hypothetical effect demands for a high accuracy of the measurements. Therefore, it is best 
to use clock transitions. 

The ytterbium atom has seven stable isotopes, two of them have non-zero nuclear spin.
The choice of isotope depends on whether the hyperfine interaction is needed to induce the transition. This is an issue for the 
clock transition 1 - 2 which is the transition between states of zero total angular momentum. It is forbidden in
the absence of hyperfine structure (see, e.g.~\cite{quenching1}) or of external field. For that reason the $^{171}$Yb isotope
which has nuclear spin $I=1/2$ is usually used for the clock. 
In the context of the present study, we do not consider odd isotopes, since in the states 4 and 5 they only 
possess $F_z\ne 0$ sub-states with very large linear (electronic) Zeeman effect.

We consider instead isotopes with zero nuclear spin (bosonic Yb), noting that it has recently been shown that high clock accuracy can also be reached with such isotopes and the 
$^1$S$_0$ - $^3$P$^{\rm o}_0$ transition in the strontium system~\cite{Rad16,Takano2017,Origlia}. 

Since we need at least four isotopes, there are the following possibilities: 
(1) Use only M2 clock transitions (transitions 1 - 4 and 1 - 5) in even isotopes.
(2) Use magnetic-field induced spectroscopy of the $^1$S$_0$ - $^3$P$_0$ transition in the four even isotopes with 
no nuclear spin.
Use either of the two M2 transitions as the second clock transition, in the same isotopes. 

\section{Comparison with other calculations}

\label{other}
During completion of our work a paper by Safronova {\em et al} on the similar subject 
appeared~\cite{SafronovaYb}.  
The authors consider another clock transition in Yb, between states 2 and 5, and suggest it for a ${\dot \alpha}$ search. 
There is some overlap between their work and the present one and
generally reasonably good agreement between overlapping results. E.g., the sensitivity of state 5 to variation
of $\alpha$  is in very good agreement. There are some differences too,
in particular in the values of the transition amplitudes and in the lifetime of state 5. 
A detailed comparison between the theoretical approaches is not possible, since not all details are reported 
in \cite{SafronovaYb}. Some preliminary comments are as follows. 

Ref.\,\cite{SafronovaYb} claims the finding of the clock transition with the highest sensitivity to the variation of $\alpha$. 
Indeed, the value of the enhancement factor is large, 
$K = 2 \Delta q/\omega_{25} = -16$ (see Table~\ref{t:alpha} for the numbers)
and this is due to small value of $\omega_{25}$. 

We note that comparing only the enhancement factors for different atomic transitions 
may lead to wrong conclusions because
the enhancement factor and the fractional measurement uncertainty of the frequency $u(\omega)/\omega$ are equally important.
This has been discussed above.
As an example,  consider the largest known enhancement factor $K \sim 10^8$ for a transition in
Dy~\cite{Dyq,Leefer}. The limit on the time-variation of $\alpha$ obtained with the use of this
transition ($ \sim 10^{-17} \ {\rm yr}^{-1}$ \cite{Leefer}) is not stronger than those obtained in systems 
with $K \sim 1$~\cite{Rosenband,Godun}. This is because the relative uncertainty 
$u(\omega)/\omega$, is also large in Dy.

When possible, including the case of the 1 - 5 and 2 - 5 transitions in Yb, it is more
instructive to 
compare the ratios $2\Delta q/u(\omega)$, as may be inferred from eq.\,(\ref{eq:figure_of_merit}). 
$\Delta q$ is similar
for both transitions ($-44\,290$~cm$^{-1}$ for the 1 - 5 transition and $-44\,290-2\,714=$
$-47\,004$~cm$^{-1}$ for the 2 - 5 transition, see Table~\ref{t:alpha}). The uncertainty $u(\omega)$
is also likely to be 
similar, because both transitions use state 5, which is the more
likely source of dominant uncertainty, due to the state's complicated structure.
Therefore, both transitions will probably have a similar figure of merit for 
testing $\dot \alpha$. 

In our opinion, the 1 - 5 transition has the advantage of being a transition from the ground state, which is experimentally simpler.

The good agreement for the $\alpha$ sensitivity between the present calculation and ref.\,\cite{SafronovaYb} is due to the fact that it is not sensitive to the incompleteness of the basis. Indeed, the relativistic energy shift of a single-electron basis state can be approximated by the formula~\cite{Qshift}
\begin{equation}
\Delta E_{\nu} \approx -\frac{1}{2\nu^3}\left(Z \alpha\right)^2\left(\frac{1}{j+1/2} - C\right),
\label{eq:q1}
\end{equation}
where $\nu$ is the effective principal quantum number ($E = -1/2\nu^2$), $j$ is the total angular momentum 
of the state and $C$ is a fitting parameter to simulate the many-body effects ($C \approx 0.6$). High basis 
states (large $\nu$) have small relativistic energy shift and contribute very little to the relativistic energy 
shifts (parameters $q$ in eq.\,(\ref{eq:q})) of the low-lying many-electron states. This can be further illustrated 
by a simple estimate used in Ref.~\cite{SafronovaYb}. If we use the relativistic energy shifts calculated earlier
for Yb$^+$~\cite{Yb+q} and the experimental energy interval between states 2 and 5, we get the correct value for
the enhancement factor $K \sim -15$ without any new calculations for the neutral Yb. 

The calculations in Ref.~\cite{SafronovaYb} 
for state 5 are performed 
with the use of the standard CI technique for 16 external electrons. 
Full-scale CI calculations in this case are not possible, 
because the CI matrix would be too large.
In ref.~\cite{SafronovaYb} the problem is dealt with by drastically cutting the basis, leaving just two or 
three single-electron states in each partial wave up to $g$-wave. 
However, 
we have shown that
up to twenty states in each partial wave are needed for basis saturation
\cite{Kramida}.
The significant cut of the basis leads to poor accuracy of the calculations. Ref.~\cite{SafronovaYb}
admits that the energies are not reproduced well in the calculations (no numbers are given).
As a result of their approximations, the accuracy for the transition amplitudes may also be poor.

\section{Feasibility considerations}

\subsection{Light shifts}

\subsubsection{Lattice light shifts}

\label{s:lightshifts}

\begin{figure}[tb]
\epsfig{figure=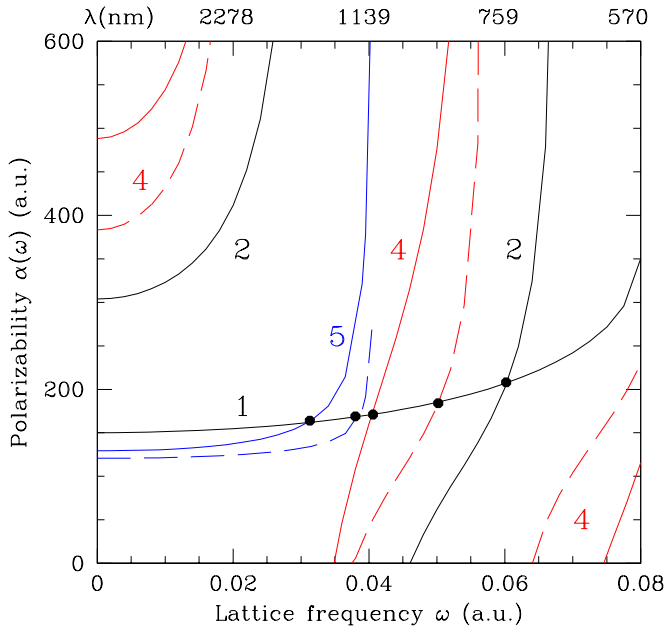,scale=0.45}
\caption{Dynamic polarizabilities of clock states 1 (black), 2 (black), 4 (red)  and 5 (blue). For states 4  and 5 
it is assumed that the lattice laser polarization is linear and only states with $J_z=0$ are considered.
The solid line corresponds to laser polarization parallel to the magnetic field ($s^2=1$),
and the dashed line to the polarization perpendicular to the magnetic field ($s^2=0$).
Filled 
circles at line crossings indicate polarizabilities at magic frequencies 
(see Tab.\,\ref{t:lambda_M}).
For state 5 the polarizabilities at $\omega>0.04$ a.u. are not shown.}
\label{f:wn}
\end{figure}

Transitions involving states with nonzero electronic angular momentum are currently not employed for clock 
applications. Nevertheless, the first demonstration of spectroscopy of cold atoms in a lattice in the Lamb-Dicke regime 
was in fact performed on such a transition, $^1\hbox{\rm S}_0\rightarrow$ $^3$P$_1\,(J_z=0)$ in bosonic $^{88}$Sr~\cite{Ido2003}. 
The observed dependence of the light shifts on the polarization of the lattice wave was considered at the time to be an issue that 
would impede achieving ultrahigh accuracy and therefore this approach was not pursued further~\cite{Derevianko2011}. 
Work on lattice clocks has focused instead on $^1\hbox{\rm S}_0 \rightarrow ^3$P$_0$ transitions, in several atomic species. 
However, one can argue that the issues arising from $J\ne0$ clock states have not yet been fully explored. 
Here we propose an approach to control
the light shifts. 

Given the $J_z$-dependence of the light-atom interaction energy \cite{Derevianko2011}, we note that it is in principle possible 
to null  the vector light shift, the tensor light shift and the first-order Zeeman shift and any static electric quadrupole shift by 
averaging over the 5 transitions 
$J_{z,b}=0\rightarrow J_{z,a}=0,\pm1,\pm2$. This nulling is independent of the polarisation state 
of the lattice field.
The magic wavelength is then determined by the vanishing of the difference of the scalar polariziabilities. 
Such a procedure would have to null rather large individual
frequency shifts, therefore we consider only a single transition, to the $J_{z,a}=0$ - state.

To second order in the electric field amplitude of the lattice field, the light shift of a transition $J_b=0\rightarrow J_a\,(J_{z,a})$ is determined by the polarizability difference~\cite{Derevianko2011}
\begin{eqnarray}
\label{eq:differential polarizability}
&&\Delta\alpha_{ab}(\omega_L)=
\alpha^{\rm S}_{a}(\omega_L) +    \label{eq:clock shift} \\
&&{1\over2}(3|\hat{\epsilon}\cdot\hat{B}|^2-1){3J_{z,a}^2-J_a(J_a+1)\over J_a(2J_a-1)}\alpha^{\rm T}_{a}(\omega_L)-
\alpha^{\rm S}_{b}(\omega_L)\  \nonumber .
\end{eqnarray}
Here, $\hat{\epsilon}$ is the polarization vector of the lattice laser. It is complex if the polarization has some degree of ellipticity. $\hat{B}$ is the direction of the small magnetic field applied to split the transition, here into the five Zeeman components  $J_{z,a}=0,\pm1,\pm2$. Note that the above expression is simplified compared to the general expression \cite{Derevianko2011} because the contribution of the vector polarizability is omitted. This is correct for an upper state with $J_{z,a}=0$ or if the lattice wavevector is perpendicular to the magnetic field or if the lattice is linearly polarized.

Thus, for a given lattice frequency $\omega_L$ and a linearly polarized lattice, the polarizability difference depends on the experimentally adjustable parameter $s=\hat{\epsilon}\cdot\hat{B}$, the relative orientation between the lattice polarization and the quantization direction. As a consequence, there is no unique magic wavelength.

Hara et al \cite{Hara2014} have performed a detailed experimental study of the light shift induced by an optical trap at the wavelength 1070\,nm on the 1 - 4 ($J_{z,a}=0,\,1,\,2$) transitions in bosonic $^{174}$Yb. In particular, the authors discussed how to produce an "effective" magic trap for each $J_{z,a}$ state by adjusting $s$.


We now consider only the upper state $J_{z,a}=0$ and a 1D lattice. Via adjustment of $s$ it is possible to minimize (when $s=0$) or maximize (when $s=1$) the polarizability of this upper state, between the values
  $\alpha_{a,min}(\omega_L)=\alpha^{\rm S}_{a}(\omega_L)+\alpha^{\rm T}_{a}(\omega_L)/2$ and $\alpha_{a,max}(\omega_L)=\alpha^{\rm S}_{a}(\omega_L)-\alpha^{\rm T}_{a}(\omega_L)$, thus maximizing or minimizing the transition frequency. Here it is assumed that the lattice wavelength is larger than 800\,nm, so that  $\alpha^{\rm T}_{a}(\omega_L)<0$ for states 4 and 5. 
Note that $\alpha_{a,max}(\omega_L)$ are given by the red and blue solid lines  in Fig.\,\ref{f:wn} while curves for $\alpha_{a,min}(\omega_L)$ are given by the red and blue dashed lines.
The magic wavelengths are given in Table\,\ref{t:lambda_M}. 
The experimental value for the magic wavelength of the 1 - 2 transition is 759\,nm~\cite{Emagic}, in good 
agreement with our computed 757\,nm.


Ido and Katori \cite{Ido2003} demonstrated a measurement of the transition frequency of Sr as a function of lattice wave polarization angle and observed a maximum and a minimum. 
Similar measurements were reported more recently for the $^1{\rm S}_0\rightarrow$ $^3{\rm{P}}_2$ (i.e. 1 - 4) transitions of bosonic  $^{174}$Yb by Yamaguchi et al.\,\cite{Yamaguchi2010}. In a  $\lambda_L=532\,$nm optical trap, they measured the light shift both for $s=1$ and for $s=0$, nearly achieving a magic wavelength condition in the latter case. 

We suggest that the transitions should be interrogated at the particular operating points $s=\hat{\epsilon}\cdot\hat{B}=0$ or $s=\pm1$, at the respective magic wavelengths 
$\omega_{M,s}$
which null the polarizability difference,  $\Delta\alpha_{ab,s}(\omega_{M,s})=0$. These operating points have also been discussed by Westergaard et al \cite{Westergaard2011} in the context of a fermionic lattice clock.
Since the transition excitation radiation must propagate parallel to the lattice wave, and a $\pi$ transition ($J_{z,b}=0\rightarrow J_{z,a}=0$) is to be excited, a suitable geometry is (i) a magnetic field perpendicular to the lattice wave propagation and (ii) a linear lattice polarization, orthogonal ($s=0$) or parallel ($s=1$) to the magnetic field.
These operating points provide extrema of the transition frequency, i.e. a quadratic dependence on the polarization setting, which is experimentally advantageous. 

\begin{table}
	\caption{\label{t:lambda_M}
		Magic wavelengths of the transitions $J_{b}=0\rightarrow J_a=2,\,J_{z,a}=0$, for particular values of
		$s=\hat{\epsilon}\cdot\hat{B}$.
		$\alpha(\omega_{M,s})$ is the common polarizability for the state $b$ and the state $a$. $\alpha^{\rm T}_{a}$ 
		is the tensor polarizability of the upper state.}
	\begin{ruledtabular}
		\begin{tabular}{ccccccc}
			\multicolumn{1}{c}{Transition}& &
			\multicolumn{2}{c}{$\omega_{M,s}$}& 
			\multicolumn{1}{c}{$\lambda_{M,s}$} &
			\multicolumn{1}{c}{$\alpha(\omega_{M,s})$}&
			\multicolumn{1}{c}{$\alpha^{\rm T}_{a}(\omega_{M,s})$} \\
			\multicolumn{1}{c}{$b - a$} & 
			\multicolumn{1}{c}{$s^2$} & 
			\multicolumn{1}{c}{a.u.} & 
			\multicolumn{1}{c}{cm$^{-1}$} & 
			\multicolumn{1}{c}{(nm)}& 
			\multicolumn{1}{c}{(a.u.)}&
			\multicolumn{1}{c}{(a.u.)}\\
			\hline
			1 - 2  &     &  0.0602 & 13200 & 757 & 208 & 0 \\ 
			1 - 4  &  0 & 0.0502 & 11020 & 908 & 184 & -204\\
			1 - 4  &  1 & 0.0406 & 8910 & 1122& 171 &-83\\
			1 - 5  &  0 & 0.0380 & 8340 & 1200 & 169 & -124\\
			1 - 5  &  1 & 0.0313 & 6870 & 1460 & 164& -24\\
		\end{tabular}
	\end{ruledtabular}
\end{table}


The operating points will be determined by extension of the well-known procedure of determining the "true" clock frequency corresponding to zero lattice intensity \cite{Ido2003}. For a given setting of 
$s^2$, close to the maximum (1) or minimum (0) value, the clock frequency is measured as function of lattice laser intensity and as a function of detuning from the magic wavelength. This is repeated for different settings of $s^2$ and the extremum of the clock frequency is determined by a fit of expression (\ref{eq:clock shift}) to the data. This is the "true", unperturbed frequency. $s^2$ can be varied by varying the  polarisation direction or the magnetic field direction, or both. 

The sensitivity of the transition frequency to $s$ is 
\begin{equation}
\delta(\Delta\omega_ {\rm LS})=0.7\,\hbox{\rm kHz}
{\alpha^{\rm T}_{a}(\omega_{M,s})\over 1\,\hbox{\rm a.u.}}
{I_M\over10\,\hbox{\rm kW/cm}^2}\,\delta(s^2)\ ,
\end{equation}
where $I_M$ is the lattice intensity. If $\alpha^{\rm T}_{a}\simeq80\,$a.u., a $1\times10^{-18}$ fractional frequency shift is produced by a $\simeq0.1$\,mrad change in angle between $\hat{\epsilon}$ and $\hat{B}$ around the operating points $s=0,\,1$ and for the reference intensity. This value is an estimate for the desirable stability of the angle over the course of the unperturbed-frequency-determination procedure and for the desirable linearity of the variation of the angle setting. 
Tab.\,\ref{t:lambda_M} indicates that the operating points $s=1$ exhibit a moderately larger angle tolerance compared to $s=0$, due to the former's smaller $|\alpha^{\rm T}_{a}|$. 

Evidently, in order to determine the optimum operation point accurately, not only the polarization optics should allow fine setability but also the intensity of the lattice should be stable over the course of the determination. Active stabilization of the lattice wave power and propagation direction can be helpful.

We consider this approach to be realistic, i.e. the unperturbed frequency can be determined in a reasonable total measurement duration, because of the low statistical uncertainty achievable with state-of-the-art clock lasers. We expect that the total uncertainty of the clock frequency related to the lattice shift only will be within a moderate factor of that achievable in conventional lattice clocks, where polarization optimization is not required.

A discussion of the atomic hyperpolarizability goes beyond the scope of this work. It is not possible to compute 
it accurately ab initio without experimental input data~\cite{Porsev2004}. However, recent theoretical~\cite{Katori2015} 
and experimental work on the 1 - 2 transition in Yb~\cite{Brown2017} demonstrates that its effects can be precisely 
measured and controlled at the $10^{-18}$ level.

\subsubsection{Probe light shift}
\label{s:probe_lightshifts}

The situation for the new transitions may be compared to that of the conventional transitions. In Yb on the 1 - 2 transition, 
a probe shift $0.8\times10^{-18}$ and uncertainty of $3\times10{-18}$ has been reported \cite{Katori}. In Sr, the probe shift 
coefficient is $-13$\,Hz\,cm$^2$/W \cite{Baillard2007}, similar to the Yb 1 - 2 transition. A shift $0.9\times10^{-19}$ and  
an uncertainty of $0.5\times10^{-19}$ have been achieved \cite{Katori}. 

The probe shifts of the new transitions (1 - 4,\,1 - 5) are estimated at $(0.05,\,1)\times10^{-19}$ for $T_\pi=1\,$s 
interrogation time, using the coefficients given in Table\,\ref{t:trans}. These shifts are comparable to those of the 
conventional transitions and therefore we expect that a similar uncertainty, in the $10^{-19}$-range, should be 
achievable. The shifts and uncertainties can be further reduced by using longer atom interrogation times.  

\subsection{Zeeman shift}

The experimental method typically used to determine the quadratic Zeeman (QZ) shift, yields an uncertainty proportional 
to the absolute value of the coefficient.

For the conventional 1 - 2 clock transition in Yb the uncertainty reported in \cite{Katori} is $1\times10^{-17}$, where the shift coefficient is given in Tab.\,\ref{t:Z2}, $-0.06\,$Hz/G$^2$. We can also quote results on $^{87}$Sr, where on the similar transition uncertainties of $\approx1\times10^{-18}$ \cite{Falke2014,Katori} have been reported, the coefficient being 4 times larger, $-0.24\,$Hz/G$^2$. 

According to Tab.\,\ref{t:Z2}, for the 1 - 4 (1 - 5) Yb transition, the QZ shift coefficient is approximately 5\,(12) times smaller than for the 1 - 2 Yb transition. Compared to Sr, the Yb coefficients are 20 and 50 times smaller, respectively.

Thus, we expect that for the proposed transitions the uncertainty of the QZ shift  can be reduced to the low-$10^{-19}$ range.

\subsection{Cold collision shift}

An important systematic effect in lattice clocks is collisional interactions between the ultra-cold atoms. In fermionic clocks these 
are effectively suppressed by using spin-polarized atoms, so that the Pauli priniciple forbids s-wave scattering~\cite{Takamoto2006}.
For bosons, s-wave scattering is a relevant interaction, especially in a 1D lattice. The clock frequency must therefore be measured 
as a function of atom density, and the unperturbed clock frequency is determined by extrapolation. 
	
The cold collision frequency shift is proportional to \cite{Harber2002} 
\begin{equation}
\rho(a_{aa}-a_{bb}+C'(a_{ab}-a_{bb}-a_{aa}))\ ,
\end{equation}
where $\rho$ is the atomic density, $a_{ij}$ is the scattering length for the collision of an atom in state $i$ and an atom 
in state $j$, and $C'\approx0.5$ is a coefficient that depends on the excitation probability and other factors. Scattering 
lengths vary widely with mass and electronic state, and cannot be computed ab initio. The scattering length in the ground 
state, $a_{bb}$, has been measured for all Yb isotopes \cite{Kitagawa2008}. Concerning the $^3$P$_2(J_z=0)$ state, 
for even isotopes only the value $a_{aa}(^{174}{\rm Yb})=-23\,$nm \cite{Yamaguchi,Yamashita} is known so far 
(for a study of the fermion $^{171}$Yb, see \cite{Taie2016}). Thus, currently there is insufficient data for a prediction 
of the density shift of even a single bosonic isotope. No data exists related to level 5. 
	
	Nevertheless, it can be pointed out that there has recently been strong progress in the accuracy of bosonic clocks using $^{88}$Sr. Transition frequency uncertainties arising from cold collisions were measured to be equal to $11\times10^{-18}$ \cite{Takano2017} and $3\times10^{-18}$ \cite{Origlia} fractionally. These clocks used 1D lattices and the technique of photoassociation to reduce the number of atoms in multiply occupied lattice sites. Also for the $^{88}$Sr isotope not all scattering lengths are known. Therefore, no strong inference from the strontium clock performance to an Yb clock performance is possible. However, it is quite possible that some of the bosonic Yb isotopes have scattering lengths of similar size as or smaller than the ones of $^{88}$Sr. Photoassociation in Yb is standard and has also been performed on the 1 - 4 transition \cite{Taie2016}.
	
The inelastic collision rate $\gamma_{aa}$ between Yb atoms in the excited $^3$P$_2(J_z=0)$ state is $\simeq6\times10^{-11}$cm$^3$\,s$^{-1}$ and, more importantly, much lower between ground-state and $^3$P$_2(J_z=0)$ atoms, $\gamma_{ba}\simeq1\times10^{-12}$cm$^3$\,s$^{-1}$. Both were determined at $<1\,\mu$K temperature \cite{Uetake2012}. In view of the values for $^{88}$Sr \cite{Lisdat2009},   $\gamma_{aa}=(4.0\pm2.5)\times10^{-12}$cm$^3$s\,$^{-1}$ between atoms in the upper clock level $^3$P$_0$, and $\gamma_{ba}=(5.3\pm1.9)\times10^{-13}$cm$^3$s\,$^{-1}$ between in ground-state and excited-state atoms, the values for Yb do not seem problematic. Moreover, we note that the Yb values were measured in a crossed dipole trap, not in a lattice, and it has been observed for the case of Sr that lower values arise in a lattice \cite{Lisdat2009}.

\subsection{Experimental implementation}
Finally, we make a few comments on the implementation. 

High-power, continuous-wave laser sources for the required magic wavelength lattices in the near-infrared spectral range
are commercially available.

Controlling light shifts at the $10^{-18}$ level will require more effort than in the standard lattice clocks, but appears feasible.

The characterization of the systematic effects of the two new  clock transitions can profit from the possibility to study both them and the standard clock transitions in the same apparatus, admitting a change of the lattice and clock lasers. In particular, this applies to the precise measurement of the black-body shift, which is dissimilar for the three transitions.

The preparation of the Yb atoms in the optical lattice can be implemented experimentally 
with the already well-established methods. That is, first and second-stage cooling can be performed with the standard 399\,nm and 556\,nm lasers and procedures. From the second-stage MOT, the atoms are released into a 1D lattice. 
The clock lasers (507\,nm, 431\,nm) excite the upper clock state via M2 excitation, which is possible, as was recently demonstrated for the 1 - 4 transition \cite{Yamaguchi2010,Uetake2012,Kato2013}. There is no difficulty in principle for realizing clock lasers for both transitions with ultra-narrow linewidth, using existing technology. 

The de-excitation of the atoms from the upper clock 
state (necessary for measuring the excitation produced by the clock laser) is already standard for state 4 as described in the cited references.
For state 5 it could be done via 
excitation to states of the  $4f^{14}6s6d$ 
configurations (using wavelengths $\sim 600$\,nm)\cite{Jin}, which will subsequently decay in steps to the ground state. 

\section{Conclusion.}

We proposed two new clock states in bosonic Yb isotopes for use 
as clock transitions in an optical lattice clock for fundamental research.
The transitions are from the ground state to metastable states with easily accessible excitation energies of 19\,710.388~cm$^{-1}$ 
and 23\,188.518~cm$^{-1}$ and angular momentum $J=2,\,J_z=0$. 

The current main motivation to use these transitions is for a search for new physics beyond the Standard Model via the 
non-linearity of King's plot and via testing for a time drift 
or a time modulation 
of the ratio of the clock frequencies. 
It is very attractive that the time drift
and modulation measurement 
could be performed with a single clock apparatus (with suitably extended laser system), similar to what is possible with 
the Yb$^{3+}$ ion.


Both transitions have a clear potential for allowing to determine and control systematic shifts with high accuracy. 
Our analysis does not indicate clear obstacles towards reaching frequency uncertainty in the low $10^{-18}$ range. This level has already been achieved for the 1 - 2  transition at NIST. 
Thus, there are prospects of further improvements of the limits of the time variation on the fine structure constant, which 
currently stands at  
$\sim10^{-17}$ per year~\cite{Rosenband,Leefer,Godun}.

Clearly, detailed experimental tests are required for investigating the validity of the proposed transitions in practice. This is especially so for the 1 - 5 transitions, which has not been studied experimentally yet. It seems that 
some key aspects of the present proposal can be characterized on existing clock devices with modest extensions. 
A measurement of the light shift will also be able to provide data allowing an estimate of the hyperpolarizabilities.

\acknowledgments

One of us (S.S.) approached M.\,S.\,Safronova with the idea of new Yb excited states for precision measurements.
We thank her for suggesting S.\,S. to contact the other authors of this paper.
 S.\,S. acknowledges a helpful hint from A.\,G{\"o}rlitz.
This work was funded in part by the Australian Research Council and by project Schi\,431/22-1 
of the Deutsche Forschungsgemeinschaft.

\end{document}